# False positives and the challenge of testing the alien hypothesis

Searra Foote[1]❋, Pritvik Sinhadc[2,3]❋‡, Cole Mathis[2,4]*, Sara Imari Walker[1,2,4,5,6,]*

[1]School of Earth and Space Exploration, Arizona State University, Tempe Arizona, USA
[2]Beyond Center for Fundamental Concepts in Science, Arizona State University, Tempe Arizona, USA
[3]Dubai College, UAE
[4]Santa Fe Institute, Santa Fe NM, USA
[5]Blue Marble Space Institute for Science, Seattle WA USA
[6]ASU-SFI Center for Biosocial Complex Systems, Arizona State University, Tempe AZ USA

*author for correspondence: sara.i.walker@asu.edu; cole.mathis@asu.edu

❋These authors contributed equally
‡Current affiliation: California Institute of Technology (Caltech), CA, USA

## Abstract

The origin of life and the detection of alien life have historically been treated as separate scientific research problems. However, they are not strictly independent. Here, we discuss the need for a better integration of the sciences of life detection and origins of life. Framing these dual problems within the formalism of Bayesian hypothesis testing, we show via simple examples how high confidence in life detection claims require either (1) a strong prior hypothesis about the existence of life in a particular alien environment, or conversely, (2) signatures of life that are not susceptible to false positives. As a case study, we discuss the role of priors and hypothesis testing in recent results reporting potential detection of life in the Venusian atmosphere[1] and in the icy plumes of Enceladus[2]. While many current leading biosignature candidates are subject to false positives because they are not definitive of life, our analyses demonstrate why it is necessary to shift focus to candidate signatures that are definitive. This indicates a necessity to develop methods that lack false positives, by using observables for life that rely on prior hypotheses with strong theoretical and empirical support in identifying defining features of life. Abstract theories developed in pursuit of understanding universal features of life are more likely to be definitive and to apply to life-as-we-don't-know-it. In the absence of alien examples these are best validated in origin of life experiments, substantiating the need for better integration between origins of life and biosignature science research communities.

## Introduction

Confirming or refuting the existence of life beyond Earth and discovering how life can originate are two of the most significant open problems in science[3], and they are not independent. Life detection proceeds by identifying observable properties that might be uniquely assigned to life, because these properties are not expected to be produced abiotically[4]. Meanwhile, to solve the origin of life, we must be able to identify when those observables unique to life first emerge from abiotic processes – in other words, we must be able to distinguish a non-living system from a living one in order to identify when the former transitions to the latter.[5,6] In this sense current approaches to life detection and origin of life science have a common goal: to identify what features of life are universal, and when we should expect to observe them.

At present there is little direct interaction between biosignature and origin of life science, most likely because making such connections requires abstracting away from the chemistry of life as we know it in a manner that so far has not dominated thinking in the field. Consider, for example, molecular homochirality as a relevant observable[7]. In life detection research, life is considered a necessary condition to observe homochirality among a set of molecules[8]. By contrast, in origins of life research a homochiral set of molecules is predominately considered a necessary *precondition* for life[9]. That one research community should regard a property as a smoking-gun signature of life, and another as a prerequisite for it suggests that progress is to be made by bridging the divide in how we ask questions in these two fields. In this example, a unified approach requires strong theoretical and empirical reasons that transcend the details of specific chemical systems to determine why life must be homochiral or conversely, why homochirality can exist outside of life.

The current situation in the mission-focused life detection and experimental origin of life communities is analogous to the situation in observational cosmology and experimental particle physics prior to the development of the Standard Model of particle physics[10]. In those fields, the realization that the early universe would have been entirely governed by the types of interactions studied in particle accelerators meant that new insights into the large-scale structure of the universe could be gleaned by conducting experiments here on Earth, and similarly that observations of the large-scale structure of the universe-imposed constraints on what to expect in high energy experiments. To make these connections, predictive theories were necessary. This realization led to a unification between those fields that spurred significant progress and large-scale international collaboration, dramatically improving the predictive capabilities and explanatory power of models in both fields, thus improving our understanding of the universe in which we live[11].

In what follows, we illustrate why life detection and origin of life studies might be poised for a similar integration if appropriate theory directed at unifying these problems is developed. First, we review the basics of Bayesian reasoning as it relates to life detection. We then motivate our arguments within the formalism of Bayesian hypothesis testing to illustrate how life detection and origin of life research are necessarily mutually informative. We demonstrate how Bayesian methods clearly articulate the trade-off between having a well-constrained and high probability prior for life (e.g., via a known mechanism for the origin of life), or a smoking-gun biosignature (e.g., by formalization of what we mean when we say something is 'life'). We discuss how the trade-off between a high-confidence signature and high-confidence prior likelihood have manifested in recent life detection claims, and we give examples of how to avoid false positive signatures. Finally, we discuss the types of data that can help constrain the prior hypotheses about the origin of life on other worlds.

## Bayesian Reasoning in Life Detection

A majority of efforts in remote and in situ life detection have so far focused on identifying signatures of 'life-as-we-know-it', that is, features such as molecules, morphology, or isotope ratios like those produced by life on the Earth[12]. This is a widely adopted approach because it is perceived as low risk given our poor understanding of the broader set of phenomena that life could be beyond our observations of the specific instantiation of life on Earth. However, this attempt at pragmatism has led to a situation where the most popular candidate biosignatures are subject to abiotic false positives, examples include atmospheric $O_2$ production, or isotopic fractionation in the rock record. These and other popular candidate signatures are subject to false positives because they can be produced by abiotic processes under some contexts and therefore are not definitive signatures of life.[13] Identifying these signatures may be interesting on its own, or it could motivate follow up investigation to determine whether the signature was indeed generated by life. However, the information required to evaluate the biogenicity of such signatures involves detailed knowledge of the planetary environment and the potential alien life that generated the signal. At present, for most planetary environments we do not have sufficient data, nor models, to make these inferences. It is possible life on other worlds always shares the chemical features present in life on Earth, which could justify searching only for these signatures, but there are no compelling arguments that suggest life everywhere should necessarily adopt the contingent features evolved on Earth. This necessitates a prior hypothesis for the likelihood of a particular kind of living process emerging in an alien environment – namely the same one as emerged on Earth - to evaluate a biosignature (e.g., oxygenic photosynthesis in the case of atmospheric $O_2$ as a biosignature). Abstracting away from the chemical features that may be unique to life on Earth to build more general theories for life is therefore important if we are to search for life as we don't know it, and if we want to increase confidence in detection even of life as we know it by avoiding false positives.

Due to the uncertainty associated with detecting features of alien worlds, astrobiologists are now aiming to quantify the (un)certainty in their detection of specific features or processes using probabilistic methods that weight the strength of their evidence.[14,15] This reasoning, weighing alternative hypotheses to explain a given set of observations and accounting for their prior likelihoods, is often captured formally by a Bayesian approach. Bayes theorem is considered to represent key features of the scientific process[16] (see also e.g., [17–19] for debate on its application to codifying science). This is because it gives the probability for a hypothesis to explain a given set of data based on observations related to that hypothesis, which can be updated by new data. For the problem of assessing the hypothesis of "life" given a relevant biosignature observable, Bayes theorem takes the form:

$$P(Life|Obs) = \frac{P(Obs|Life)P(Life)}{P(Obs)}. \qquad (1)$$

Bayes' theorem (Eq. 1) captures how confidence in a hypothesis (Life) based on some new observation (Obs) should be equal to the confidence that the observation (Obs) would be generated if our hypothesis (Life) is true, multiplied by our prior confidence in the hypothesis (Life) and divided by our prior confidence that the observable would result regardless of whether our hypothesis is true or not. This decomposition, even without adding any details about life or its observables, already contains implicitly a theory of living systems that is important to make explicit - specifically it assumes life is either present or absent, and there is nothing in between. This is a simplifying assumption that is motivated by the fact that nascent biospheres or abiotic worlds undergoing the transition to life may not have detectable differences from completely abiotic worlds. This assumption could be relaxed, because life/not-life is not necessarily a binary categorization and this categorization will ultimately depend on the theoretical framework that ultimately allows us to explain what life is.[20–22]

Our focus here is to explicate how Bayes theorem illuminates the need for having either a smoking-gun biosignature, or in the case of biosignatures subject to false positives strong support for the prior existence of life in that environment. We motivate this argument using a simple coin toss example.

**Flipping a Biased Coin.** Imagine you have two coins. One is a fair coin with a side for heads and a side for tails, while the other is a biased coin where both sides are tails (see Figure 1). You randomly choose one of the coins and you do not know which it is, so you decide to guess by flipping it and infer which coin you have picked up. The outcome can be heads or tails, but because you are uncertain about which coin you have, the probabilities are conditional on the likelihood you have selected either coin: a tails result from the fair coin with chance $P(T|F)$, a heads result from the fair coin $P(H|F)$, a tails result from the biased coin $P(T|B)$, and a heads result from the biased coin $P(H|B)$. Let's say you observe tails when you flip it. Which coin do you have? The answer to this question can be quantifiably assessed with Bayes theorem in the following form:

$$P(T) = \frac{P(T|F)P(F)}{P(T)}. \quad (2)$$

By definition $P(T|F) = 0.5$ because we expect a fair coin to yield a fifty-fifty chance of tails. $P(F) = 0.5$ because our assumption is that the coin you have was randomly selected (50% probability). To calculate the prior likelihood of the tails result needed for the denominator, we can decompose it into the conditional probabilities for getting tails from either coin:

$$P(T) = P(T|B)P(B) + P(T|F)P(F),$$
$$P(T) = (1.0 * 0.5) + (0.5 * 0.5) = 0.75 \quad (3)$$

Using Equation 2, this yields the result $P(F|T) = 0.5 * 0.5/0.75 = 1/3$ - that is, if you flip your coin and observe tails, you can have a confidence of approximately 33% chance you have picked up the fair coin. Meanwhile if you had observed heads you know with 100% confidence you have the fair coin in hand (because the biased coin has no side labeled heads). In Figure 1 we show how this calculation would change if the biased coin did not always produce tails (e.g. different values for $P(T|B)$) or if you had a very different prior expectation that you had the biased coin (e.g. $P(B)$). In the case of this simple example, we are given these probabilities ahead of time, but in the case of life detection these values are unconstrained (we do not know them *a priori*, in fact it is what we aim to learn by searching for life).

In Bayesian analysis probabilities *do not* represent the frequency of specific outcomes after repeated observation or experiment. Instead, they represent the confidence in that outcome being an accurate model of reality. This might seem like a subtle distinction, because in many cases it does not matter. For example, if you are told a fair coin flip has a fifty percent chance of landing heads up, it doesn't matter whether the information is about the average over a series of observations (frequentist interpretation) or the confidence in the outcome of single flip yielding heads (Bayesian interpretation).

However, for many other types of claim, like one claiming there is a fifty percent probability a planet hosts alien life, we are not making a statement that out of many previous observations of that planet half of the time it has no life – instead we are quantifying our uncertainty about a presumably objective, observational feature of that planet. We assume there are some objective features of the system that we simply do not have access to, but which would reduce our uncertainty to zero if we did. Without observing those features directly, the best we can do is guess and to quantify the confidence we have in that guess.

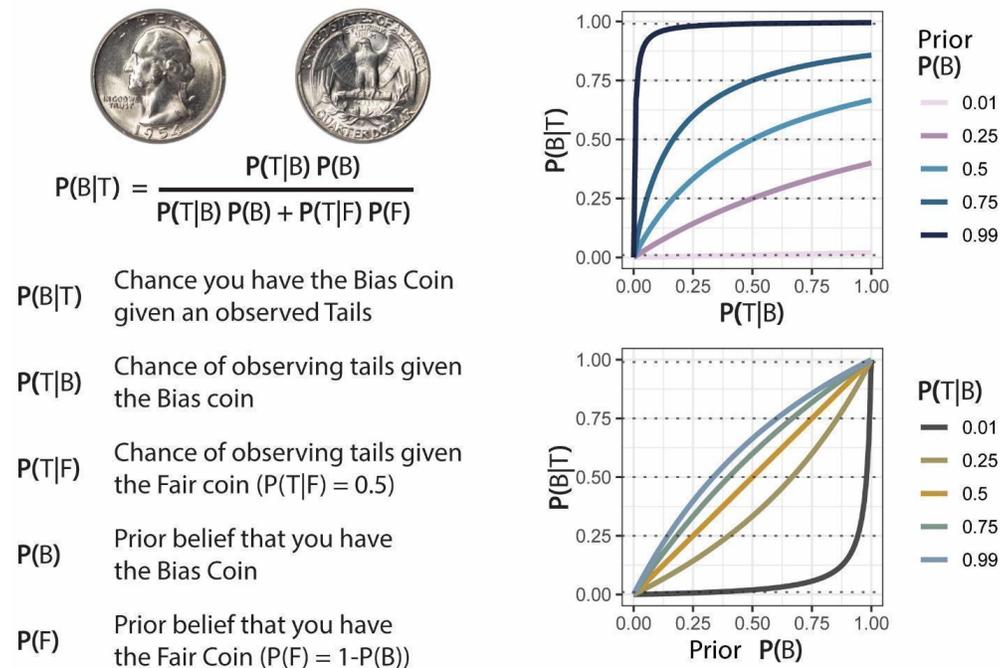

Figure 1. Coin Toss Example. There are two coins. One is fair (such that the probability that it shows heads or tails are equal). The other is biased such that it is weighted to yield tails result more often, as determined by the probability P(T|B). If you observe a tails outcome you can calculate the probability you have the biased coin using Bayes equation. The top right shows how this probability changes for different biases in the coin, and the bottom right shows how this probability changes for different prior probabilities corresponding to the probability you selected the bias coin to begin with.

In this coin toss example, we are discriminating between just two hypotheses – whether we have in hand a fair coin or a biased coin. This can be roughly analogized to the problem of detecting life, where we want to determine whether we have observed a planet that has a biosphere or does not (again we are making a naive assumption this is a binary categorization). Ambiguity in our result arises because we did not *a priori* assign a higher likelihood to one of our hypotheses over the other, and because the data we happened to observe cannot discriminate between the two hypotheses.

In life detection, when we encounter non-discriminatory biosignature data we often discuss the possibility of "false positives" – features that could be misinterpreted as signs of life when they are in fact produced abiotically. An example is the possibility of an abiotic world producing abundant atmospheric $O_2$, due to the presence of liquid water on its surface with photolysis driven by its parent star[23]. This is a false positive scenario because without the knowledge of this abiotic formation mechanism, we might otherwise have assumed $O_2$ is a biosignature produced by living, photosynthetic activity. And indeed, early in exoplanet science atmospheric $O_2$ was regarded as a "smoking gun" signature of life (meaning it was sure-fire evidence for life). If we had not identified abiotic mechanisms for $O_2$ production, we might still regard $O_2$ as a definitive biosignature. For biosignatures subject to false positives, because they are not derived from a theory for differentiating life from non-life, it is impossible to know if we have exhausted all possible abiotic mechanisms. For the case of $O_2$, even if we assume all planets with life always produce $O_2$ atmospheres (a likely gross over-estimate given that there is a large observational window in our own planets' early evolution where this is not the case), our ability to interpret it as a biosignature should still depend on the likelihood of abiotic planets producing abundant $O_2$. That is, we must know the abiotic formation mechanisms before we can even assess the likelihood of biological origin. For most current biosignature candidates, an exhaustive search over all possible abiotic formation mechanisms is not possible, either because they are not known, or because not enough of the planetary environment is known to fully specify them. Because it is possible for an abiotic source for our data, we must also know the likelihood of the emergence of life having happened in our dataset, which we currently also do not have access to. If we assess the likelihood of false positives to be high, we need a higher prior likelihood for life and tighter constraints in our confidence of our assessment.

This overly simplified $O_2$ example is in direct analogy to the 'tails' result, where we can equate inhabited planets to the biased coin (always produces tails is analogous to always produces $O_2$ atmosphere) and uninhabited worlds to the unbiased coin (sometimes produces tails is analogous to sometimes produces $O_2$ atmosphere), see Figure 2. More specifically, the rate of false positives is the probability of the observation conditioned on a non-living planet P(Obs|NL). Notice that if the probability of the false positive P(Obs|NL) is zero, then P(Life|Obs) is equal to one, independent of the prior probability of life (P(Life)). This is the key point. If we have biosignatures which are susceptible to any false positives, we require high prior expectation for life producing the signature, which must be larger if the false positive rate is higher. If we know that a certain observation can only be made by living systems in all situations, we do not need to reason about the prior probability of life emerging in an alien environment to generate it, its detection is sufficient to make a life detection claim.

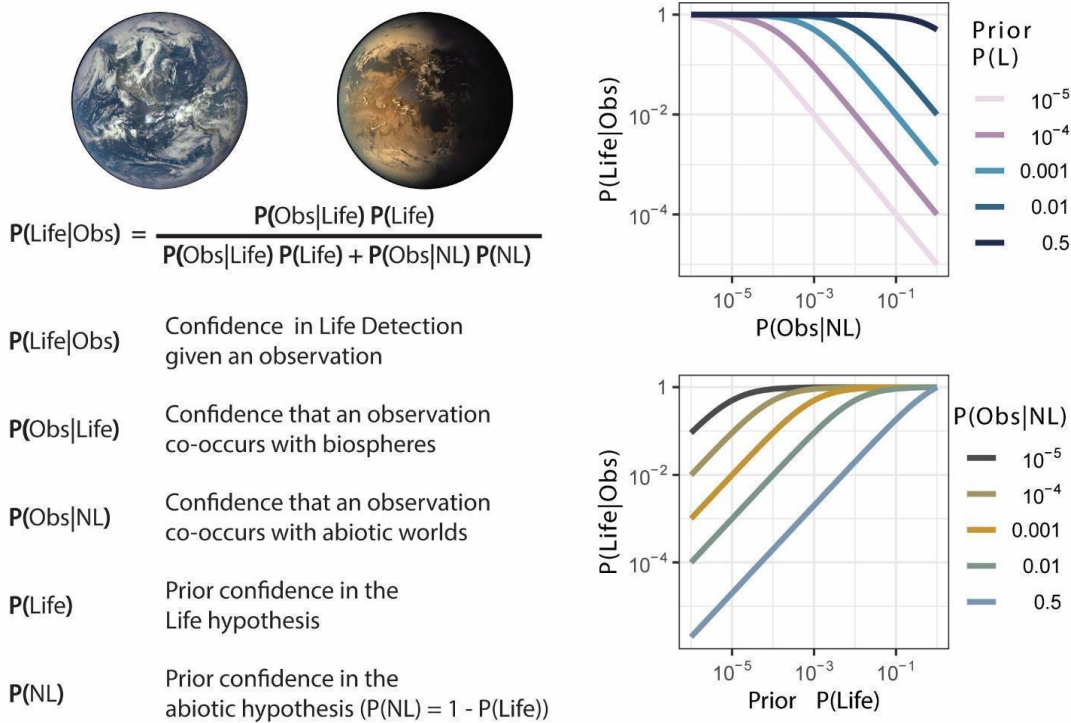

Figure 2. The complex problem of detecting biospheres is simplified to the analogous situation of determining whether a coin is fair or biased. If instead of testing the hypothesis "I have the biased coin," we want to test the hypothesis that "this planet has a biosphere," we can formulate it using a similar approach. Again, it depends on the probability a non-living planet produces an observable, P(Obs|NL) (top right), and the prior confidence we have that the planet hosts a biosphere, P(Life) (bottom right). In both graphs we have assumed that P(Obs|Life) = 1.0 such that our observation is expected to be produced by all biospheres. P(Obs|NL) is the false positive rate— how often do abiotic worlds produce the observable. If P(Obs|NL) is low (on a logarithmic scale) we do not need a strong prior for the emergence of life in that environment to have significant confidence in our life detection claims.

The coin toss example is meant to simply illustrate one key point: avoiding false positives requires a strong prior hypothesis for why life should be expected to exist in a particular environment and that the origin of life happened in that environment (i.e., you are confident which coin you are given). This is true even if one accounts for context that decreases the likelihood of false positives to anything other than zero – if we do not rule out false positives entirely, we cannot avoid this issue. The alternative is then that life detection should proceed with biosignatures that are not subject to false positives (i.e., the coin landed on heads).

**Life Detection on Venus and Enceladus.**
In 2021 there were two potential claims of life detection on two different targets in the solar system, published separately in the same journal just a few months apart. One made international headlines; the other was read mostly by specialists in the field. Why the difference? The two potential claims were related to an assessment of a biological explanation for methane in the plume of Enceladus and to the discovery of phosphine on Venus. Both methane and phosphine are candidate biosignatures subject to false positives.

Phosphine has been proposed as a biosignature candidate in the context of exoplanet science because it is a remotely detectable volatile gas believed to be uniquely produced in some environments only by life[24]. Its detection in the upper atmosphere of Venus therefore raised the possibility of a biological origin. Subsequently the detection of Phosphine has been debated, and we refer the reader to[1,25–27] for more details on that debate. Our focus here is on the reasoning associated to claims of alien life detection. Therefore, we assume a positive detection of phosphine for purposes of illustration with our points holding in general, and independently of whether this specific case continues to be one of debate in the literature.

Because phosphine is producible abiotically, attributing its presence on Venus to biology requires determining the likelihood it could be produced abiotically. To approach this, Greaves *et al* constructed a series of abiotic models based on thermodynamic calculations and photochemistry[1]. For most mechanisms they tested, the expected production rate of phosphine was much lower than what was inferred from the observation, often by several orders of magnitude. This analysis did not test specific hypotheses related to the possible biological processes that could have produced this signal, they only attempted to rule out possible abiotic mechanisms. Ruling out all possible abiotic mechanisms is a daunting task, even for relatively well studied objects in the solar system. Nonetheless, in the case of phosphine on Venus many abiotic mechanisms could be ruled out, yielding the proposal that life might be the more likely explanation. In Bayesian hypothesis testing it is straightforward to see why this is a premature conclusion. The focus on minimizing the false positive rate of P(Obs|NL), without a parallel attempt to rigorously evaluate the prior P(Life) means we are left with a conditional likelihood in the absence of context for assessing how it impacts our conclusion. This is problematic because we know that P(Obs|NL) is non-zero, abiotic processes in the lab can produce phosphine[28] and the atmosphere of Venus still has many properties subject to scientific debate. Ultimately this meant that the study generated significant claims and debate about the source of phosphine on Venus without being able to moderate the strength of those claims (on either side – in favor or not of the alien hypothesis) based on what we might reasonably constrain about the likelihood of a biosphere on Venus[29]. In fact, in a different study the same authors also noted that "the hypothesis that phosphine is produced by life cannot *a priori* be favored over the hypothesis of unknown photochemistry or unknown atmospheric chemistry."[30]

This should be contrasted with the less controversial life detection claims on Enceladus[2]. Affholder *et al.*, estimated the escape rates of methane in Enceladus' plumes, and compared them to the measurements made by the Cassini Mission[2]. Their analysis tested two different hypotheses – (1) methane was produced via abiotic serpentinization or (2) methane was produced via biological methanogenesis. Of those two explanations tested they found that the escape rates have the highest likelihood under the hypothesis of methanogenesis conditioned on the prior hypothesis that the probability of an origin of life event in the environment of Enceladus is high. If the prior probability of life emerging is low, then they claim Enceladus could be classified as uninhabited with an unknown source of methane. The example presented by Affholder *et al* is important because it illustrates in a very clear and simple manner how evaluating the 'alien hypothesis' for biosignatures subject to false positives depends critically on both our understanding of possible abiotic processes and on the prior likelihood of life's emergence in that environment. The focus on determining the conditions under which P(Obs|Life) is large enough to interpret the explanation for methane as deriving from life with high confidence, could only be done by simultaneously evaluating P(Obs|NL) and determining the dependency on P(Life). This meant that the study produced no significant claims about the source of methane on Enceladus, only a strong indication that *if* a mechanism for the emergence of life and evolution of methanogenesis could be placed in the environment of Enceladus, then we could interpret the methane as biological.

## Ladders versus Conditional Probabilities

The modern history of astrobiology is dotted with numerous claims of alien life detection. Yet we do not all agree we have made first contact, because after their initial announcement all of these were subsequently found to be ambiguous, similar to the case of phosphine on Venus. It is therefore understandable that space agencies such as NASA, which have a vested interest in high-confidence life detection and accurate reporting to the public, should take on the task of attempting more standardized methods for assessing claims of the detection of alien life. Recently these have come in the form of "ladders," meant to rank the priority of assessing specific features of a life detection claim in a fixed set of steps[13,31,32]. Most recently Green *et al* suggested a standardized scale they call the Confidence of Life Detection (CoLD) Scale[31], see Table 1. The scale is built on a sequence of evidentiary thresholds that must be met before a claim related to life detection can be made[31]. A community wide effort to establish a similar scale proposes five criteria for evaluating life detection claims, broken into two high level categories, which can be addressed non-sequentially[32], see Table 2. The first two criteria- categorized as level I - are technical questions related to (1) identifying the signal of interest and (2) evaluating the source, strength, and possible confounding factors in the data analysis (currently this is where the debate of phosphine on Venus is most intense as now the signal itself is debated). These are separated as level I criteria because they do not reflect conceptual concerns unique to astrobiology: detecting anything with confidence on other worlds is a technical feat in any field of planetary science. The level II criteria include (3) ruling out abiotic explanations, (4) ruling in biological explanations, and (5) looking for alternative lines of evidence[32].

| CoLD scale Level | Corresponding Measurement indicator |
| --- | --- |
| Level 1 | Detection of a signal known to result from a biological activity |
| Level 2 | Contamination ruled out |
| Level 3 | Demonstration or prediction of biological production of signal in the environment of detection |
| Level 4 | All known non-biological sources of signal shown to be implausible in that environment |
| Level 5 | Additional, independent signal from biology detected |
| Level 6 | Future observations that rule out alternative hypotheses proposed after original announcement |
| Level 7 | Independent, follow-up observations of predicted biological behavior in the environment |

Table 1: CoLD (Confidence of Life Detection) Scale proposed by Green *et al* 2021 [31].

| Level | Standards of Evidence Criteria | Question |
| --- | --- | --- |
| Level I | 1 | Have you identified the signal of interest? |
|  | 2 | What confidence do you have in the source, strength and possible confounding factors in the data analysis? |
| Level II | 3 | Have you ruled out abiotic explanations? |

| | 4 | Can you rule in biological explanations? |
|---|---|---|
| | 5 | Can you identify alternative lines of evidence? |

TABLE 2: Standards of Evidence Life Detection scale, produced by a community wide effort [32].

While these efforts can be productive for organizing community conversations about the future of the field, they do not represent the flow of scientific progress nor do they necessarily reflect the coupled conditional probabilities inherent in any alien life detection claim that aims to disentangle abiotic and biological explanations. For one, these scale-based approaches typically separate assessment of the life hypothesis from the assessment of abiotic explanations. However, as we have already emphasized, these probabilities are conditionally dependent on one another– they are not independent hypotheses. Any assessment of a signal that is known to be produced by living and non-living systems must also rely on the prior probability for life to exist in that environment. This is problematic because it means that evaluating the strength of a biosignature requires strong theoretical support not explicitly addressed in the scale format. Any evaluation of the prior probability of life will be primarily theoretical, not empirical. This is because it will depend on counterfactual statements in the form of if/then statements: for example, if there is a geochemically sustained chemical disequilibrium, life is more likely. Counterfactual statements are inherently theoretical because they are claims about the world as it could/might be, not as it is, which requires having built a theory of the world in which to test scenarios that could play out. The strength of the theories used to evaluate the prior probability of life is not included in the evaluation of the overall life detection claim. Additionally, the scales make an explicit assumption any life detection claim will always be subject to false positives. While historically this is the case, it need not be so. We can avoid the problem of false positives if we build general theories that allow us to identify objectively what life is (assuming life is an objective category, but if it is not, it is unclear what we should be looking for[33]), or if we target more robust features such as signs of technology that are not subject to false positives (see Avoiding False Positives below).

The astrobiology community is attempting to make concrete steps in a positive direction, based on past lessons learned and an increasing amount of data on proposed biosignatures. This includes development of the Life Detection Knowledge Database (LDK). LDK organizes entries for biosignatures to assess which of these would be of most value to the astrobiology community and includes an aim of streamlining the process for scientists to generate science traceability matrices for mission proposals and for funding agencies to review them. Each biosignature in the current LDK includes metadata on the Biological Prevalence, Biological Feature Strength, Abiotic Prevalence, and Abiotic Feature Strength. Prevalence is defined as an expression of the likelihood that a particular feature will be produced by a biological source, while the Biological Feature strength is defined as the prominence of a feature that provides evidence for the existence of past or extant life. Abiotic prevalence and Abiotic Feature Strength are defined similarly for abiotic processes. However, it is currently impossible to integrate these entries into quantitative estimates for the conditional probabilities for them to occur – most of the biosignatures input into the LDK were not designed with probabilistic assessment in mind, nor do they explicitly connect to more generalizable abstract principles about the nature of life to which we might assign such probabilities.

Some of the conceptual challenges arise because the models and data used to make arguments of relevance to life detection are often from completely different disciplinary domains, making the resulting conclusions difficult to compare directly. But the main problem is that the studies which are used to motivate each argument are themselves not translated into probabilistic statements, based on the current uncertainty of the models and data. The LDK makes clear how integrative, and quantitative approaches to biosignatures will not be possible until astrobiologists commit to making quantitative claims about biosignatures, which itself requires more sophisticated approaches to modeling and theory development. That is, the existence of the LDK plays an important role because it allows us to precisely articulate these kinds of limitations, motivating adoption of theoretically derived biosignatures that allow quantitative assessment of the likelihood of life from a given set of observations. Such quantitative approaches are common in other disciplines but have not been cultivated yet in astrobiology, although increasing effort is being made. The LDK could serve as a point of organization in aggregating the data and models needed to quantitatively evaluate biosignatures, which will provide estimates of P(Obs|Life) and P(Obs|NL) and how they might vary across environmental conditions. To understand how those estimates will influence life detection claims we also need constraints on the prior probability of Life, P(Life).

## Constraining Priors via Experiments on Earth

The actual probability for life to emerge from planetary chemistry is currently relatively unconstrained. This is because we have observed the evolutionary products of only one origin of life event in the universe, and for that event we can only wonder about it because we are the product of it and have evolved sufficient intelligence to wonder[34]. There are three sources of new data that could lead to better constraints on P(life): (1) new evidence of life earlier in Earth's history, (2) new knowledge of the emergence of life from experiments in the lab, and (3) detection of another biosphere, see Figure 3 (and see References[14,35] for discussion of how these revise Bayesian likelihoods in life detection and Reference[36] for a discussion of their relevance to origins of life).

Chen and Kipping 2018 argue the most informative data should come from a large-scale survey of Earth-like exoplanets in a search for biospheres[35]. They show, using Bayesian arguments, how such large-scale surveys will provide the data most likely to change the shape and mean of the probability distribution for the likelihood of abiogenesis per unit time, at least on Earth-like planets. This data could be useful in constraining the likelihood of the origin of life even if none of the planets surveyed had detectable biospheres. There is a good analogy of why this should be the case, which comes from particle physics. The Super Kamikande experiment has as its stated goal to observe the decay of the proton, that is, if it is allowed by the physics in our universe[37]. Super Kamikande is looking for decay events, which have so far never been observed (in analogy to alien life which is likewise not yet observed). As time passes, non-detection events still provide information and yield a larger lower bound on the decay timescale of the proton, which constrains the structure of theories in fundamental physics (by excluding those that would predict the proton to decay with a shorter lifetime than the current observational lower bound). Likewise, the more potentially inhabited planets where we observe they are not inhabited by life, or at least Earth-like life, the better bounds we can place cosmologically on the distribution of life (or Earth-life) in the universe (which in turn is related to its probability of it occurring).[37]

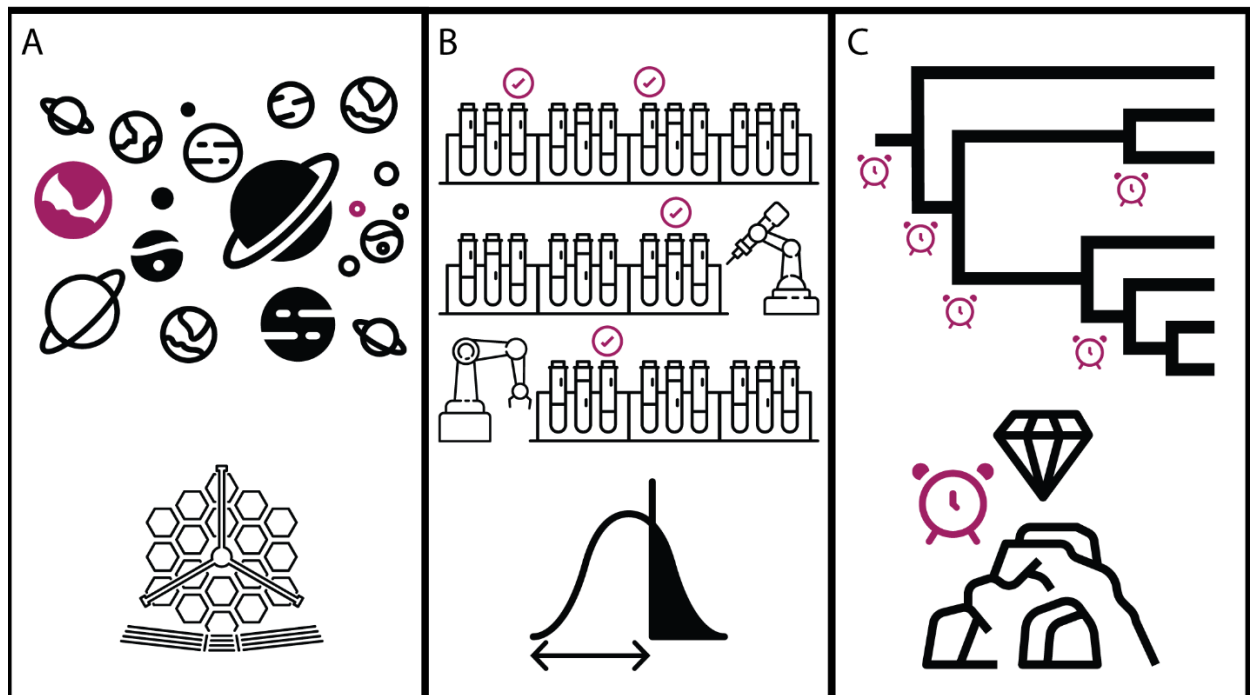

Figure 3. The three ways to estimate the prior distribution of life P(Life). A) Assessing the probability of life on other worlds via large scale surveys. B) Ensemble measurements and exploration of chemical space in search of the origin of life. C) Tighter constraints on the timing of life origin on Earth and better constraints on major transitions in Earth life.

The challenge with surveys as the solution to the prior probability problem is that the data we will acquire from exoplanets in the coming decades will be limited by the technological challenges of remotely inferring the properties of planets orbiting distant stars. These challenges, combined with the lack of clearly articulated theories of life could prevent accurate interpretation of the data. Therefore, the constraints we might get on P(Life) – even if we conduct surveys to bound the probability, are likely to be only weak bounds (by observing certainly dead worlds for example) in the foreseeable future, even considering data we are likely to acquire from recently launched (JWST) or proposed (LUVOIR/HabEx) missions[38].

To interpret exoplanet data or data from our own solar system, we need lab-based origins of life experiments on Earth that can help bound the probability of life arising across different planetary environments. Chen and Kipping included a thought experiment to understand how this could work, with experiments to essentially represent virtual planetary environments, with many replicates started under identical initial conditions[35]. Observing when these simulated planetary environments emerge de-novo origin of life events would constrain the minimum amount of time required for the emergence of life (even if no event is observed, as with Super Kamikande). Chen & Kipping 2018 framed this as a thought experiment. However, this survey must be performed in real chemical experiments, and it is possible to do so in the near term, given advances in digital chemistry that are explicitly targeted at building ensemble experiments to search chemical space for the origin of life[39–42]. These newer approaches to origins of life depart from attempts to engineer specific biochemical components under precisely controlled conditions and aim instead to solve the origin of life as a search problem in chemical space. We cannot run this search in silico because we do not know the mechanisms under which life arises[43]. Indeed, the origin of life as a chemical search problem is targeted specifically at enabling discovery of new principles underlying life, allowing us to evolve in essence alien life in the lab.[5,39,44] These approaches also present a controlled environment to develop agnostic life detection techniques which will be required to detect truly novel life forms.

Our uncertainty in the processes involved in life's emergence means the space of possible experiments is vast, and the timescale required may be much longer than the timescale of typical chemical experiments (1-7 days). To explore this space a combination of massive scale chemistry experiments (similar in cost and size to those seen in particle physics and astronomy, but not yet seen in chemistry), and theoretical developments to guide the experiments and interpret the results is necessary. If these experiments are to constrain the prior likelihood of life over all possible exoplanet environments, they will need to be developed in a completely different framework to that of typical prebiotic chemistry experiments, which have aimed to model the emergence of our biochemistry by engineering constraints and targets for synthesis based on our own planetary history. The space of possible worlds and possible biochemistries is sufficiently large to render the established approaches of prebiotic chemistry uninformative to the broader question of life detection.

The goal of such large-scale experiments should first be to understand the relationship between chemical initial conditions and the chance of living processes emerging. At present is not clear if there is a narrowly defined set of chemical conditions which reliably lead to the emergence of biochemistry, or if the same set of initial conditions can lead to a diversity of contingent possible biospheres, or if the process is convergent from a broad set of initial conditions, or if it would on rarely produce a biosphere of any description. Understanding the relationship between chemical initial conditions and the observable features of the biochemistry (or lack thereof) produced should be the top priority for the large ensemble experiments described here. These could be specific measurements[45], such as looking for evidence of increasing molecular assembly (described in the next section), or for the emergence of broader macroscale features that evolution might converge on[46,47]. That data could be used to understand how the prior probability of certain living processes depends on planetary conditions by understanding how those planetary processes set the chemical initial conditions. A key consideration in these experiments will be the timescale required to observe living processes. The timescale required for the emergence of Earth life is unknown but could be on the order of seconds to 500 million years. However, the timescales of chemical phenomena are set by the diffusion of molecules in aqueous solution, which suggests the key chemical processes should happen relatively rapidly. This leads to the question of whether we can trade time for space (volume of chemical experiments) in solving the origin of life, or if indeed there is a contingent set of steps, each of finite duration necessary for life to emerge (in which case we cannot trade time for space). The uncertainty is in the degree to which solution chemistry couples to geological processes which would introduce much longer waiting times. But for lab-based experiments this coupling can be simulated via automated changing chemical conditions. Algorithmic control of massively parallel chemical experiments could therefore dramatically reduce the timescales for an origin of life event to occur, and would allow us to extrapolate to relevant timescales on planets.

## Avoiding False Positives

The history of life detection is a history of false positives. We have already discussed a few such cases but perhaps the most classic example is the announcement by then US President Bill Clinton on the lawn of the White House of possible Martian microfossils in the Allan Hills meteorite. The hype-cycle of 'ah-ha' claims of alien life detection and subsequent realizations that the problem is much more nuanced and challenging, have led to popularization of the mantra "extraordinary claims require extraordinary evidence"[48]. However, the reasoning we have outline here, based in a Bayesian assessment of the scientific process illustrate that this is not quite right, and the mantra might be more aptly put as "extraordinary claims require extraordinary explanations" – that is, the hypotheses we come into life detection with, and the strength of their theoretical support are absolutely critical. The evidence itself may in the end be quite simple. To break the hype-cycle of false positive life detection claims, we cannot leave the hypothesis of "life is present on the planet," as one of last resort, instead we must organize the evidence required to critically evaluate our (un)certainty.

Given the plethora of false positive scenarios for current biosignatures, there is widespread assumption that biosignatures with false positives are unavoidable. This is not true, and rests on the corollary assumption that we do not know what life is. However, this logic cannot hold – if we do not know what life is, and attempt to identify it based on biosignatures that can also be produced without life, how can we possibly validate what we have found?[33]

One of the challenges in life detection is the fact that at the microscopic level living matter is subject to the exact same laws of physics and chemistry as non-living matter. Put another way– life does not violate the known laws of physics or chemistry. Therefore, many astrobiologists believe biosignatures must always have false positives. But living matter is clearly different from non-living matter, and if those differences are more than anthropic artifacts of our own perceptions (they are objective), there must be a way to measure them. If they are not, we can never hope to detect life because its distinction from non-life is not a natural category. The way to develop biosignatures that are not susceptible to false positives is to identify properties that are only possible because of life. Technosignatures provide an example: the only way we could receive a radio broadcast of the 1932 Olympic Games from another star system is if an alien biosphere observed our world and sent the signal back to us. It is not that there are no abiotic sources of radio waves, but rather the structure of that particular signal is so specific that it is exceedingly low probability to ever observe it, and its detection would imply long enough integration times that the signal would have to have been be sustained by an intelligent sender (e.g. repeated over time). The combination of non-trivial structure and high copy number is only explainable via intelligent systems[49]. If we detect sustained transmission of an identical signal encoding rich information, we do not need to reason about the probability of life forming in that star system because the most parsimonious explanation of the data is that a biosphere generated it, regardless of how likely a biosphere was to emerge in those conditions.

Assembly theory provides an example of a biosignature which avoids false positives by explicitly formalizing this intuition. Molecular assembly numbers quantify the number of operations required to construct a molecular graph and therefore tracks the specificity of the molecule in the set of all possible alternatives[45,50,51]. The key claim of the theory is that assembled objects with many steps cannot form in the absence of life, and therefore you will never observe them in high abundance abiotically. In other words, assembly theory is based on the assumption that the laws of physics do not contain the design of complex objects (they do not emerge spontaneously in the universe without living systems) and therefore if you observe multiple copies of an object with a large minimal number of steps required to produce it, it is a sign an evolutionary process having selected for that object to exist, e.g., it is a sign that a 'machine' exists that was selected to construct the object, and therefore is a sign of life. This intuition is empirically validated: high assembly number molecules in detectable abundance have been shown to only occur in living systems, and therefore detecting them on another world would constitute a biosignature not susceptible to false positives[45]. The theoretical reason is the same reason a radio wave encoding the Olympics would be a biosignature: its high specificity and reliability (reproducibility of the signal). Abiotic processes produce molecules but the production of identical copies of nontrivial structure is not possible without the information processing systems of living matter. A feature of this biosignature is that it can be tested experimentally on Earth without recreating an entire planetary surface or biosphere, and therefore it is falsifiable. If we detect high assembly molecules on another world we do not need to reason about the probability of life forming on that world because the most parsimonious explanation of the data is that a biosphere generated it, regardless of how likely a biosphere was to emerge in those conditions. There is strong theoretical support for this intuition, a mathematical theory to quantify it, and experimental tests that give the theory empirical validation[45,49,52]. This iterative loop is leading to the design of the kinds of experiments we described in the last section, and new methods for looking for life in mission-based efforts.

Life detection based on molecular assembly is not subject to false positives in the same manner as life detection using anthropocentric biosignatures (amino acids, atmospheric $O_2$) because it is designed to detect a phenomenon which only exists conditioned on the existence of an evolutionary process e.g. P(Life|Obs) = 1.0. This does not depend on the environmental context of the signal or the properties of the host star. The contextual information needed to evaluate such an observation is merely the information required to verify the signal itself. This is conceptually different from the detection of biosignatures which require contextual information to verify the signal and additional information to strengthen the claim that the biosignature is indeed derived from life, and that furthermore would not tell us any of the features of the evolutionary process and life that underlie it (i.e., how selected the chemistry is[49]).

There are suggestions that a sufficiently tuned planetary environment could produce high assembly molecules. However, this argument does not hold, because it underestimates the size of chemical space and the specificity required to produce complex objects, which if accounted for means the counter argument is essentially tantamount to intelligent design (environments are specifically fine-tuned to produce complex objects). Most would agree we would never find a cellphone (a complex object) forming from the geochemistry of Mars, without an evolutionary process (i.e. us) putting it there. Assembly theory formalizes this intuition for molecules, and allows and empirical method for determining where this probability boundary lies in chemical space (for Earth-based chemistry it is above assembly index 15[45]). Geochemical constraints and heterogenous environments may drive some aspects of chemical synthesis in surprising ways, but there will be limits to what is possible without encoded information processing systems as seen in biology, experimental approaches described above can help identify those limits, as well as chemical exploration within the solar system[53].

## Conclusion

The detection of life on another world is a critical research objective for astrobiology. To make confident claims about such a detection, the astrobiology community must begin to integrate data and perhaps more importantly theory, across the disparate disciplines and research programs that compose astrobiology. As we demonstrated, Bayesian hypothesis testing provides a simple framework within which to understand the challenges and benefits of this type of integration. Specifically, this framework illustrates the dependence of our claims on both the putative biosignature signal and the prior hypothesis about life's origin in each environment. It also provides a clear methodology for moderating life detection claims based on the available evidence, without pre-specifying the individual scientific steps required to make the claim which might reasonably differ based on different observational targets, measurements, and its significance. However, we do not mean to prescribe Bayesian Hypothesis testing as the only acceptable quantitative framework for life detection – there may be other alternatives which are better suited to the problem and the needs of the community. Rather we use Bayesian Hypothesis testing as a useful tool to illustrate the co-dependency of origins of life studies and life detection within astrobiology, and the need to move towards more abstract ideas about the nature of life that might allow us to unify the two to assess putative life detection discoveries more confidentially.

If we can only use biosignatures which are susceptible to false positives we need strong constraints on the prior probability of life on other worlds. If we have in hand theoretically motivated and empirically supported biosignatures that are robust to false positives, we can make strong claims about life detection even if we do not have a strong hypothesis about life's origin. The most direct way to identify biosignatures which are robust to false positives, and to place constraints on the prior probability of life is via the rapid exploration of abiotic chemistry– which can be done via experiments on Earth if guided by appropriate theory. Therefore, the goals of life detection represent a focal point between origins of life research and the search for alien life, and progress in both will require theories that can capture the nature of living processes, and experiments which can both expose that nature and test the theories which describe it by evolving alien life in the lab.

## Acknowledgements

This work was supported by NASA Grant 80NSSC18K1140 (to S.I.W.).  CM would like to thank support from NASA through the Postdoctoral Fellowship Program. The views and conclusions contained in this document are those of the authors and should not be interpreted as representing the official policies, either expressed or implied, of NASA. We wish to thank Leroy Cronin for many fruitful conversations inspiring our discussion of origins of life experiments and life detection, and Harrison B. Smith for productive conversations and critical feedback on an earlier version of this manuscript.

## Author Contributions

SF and PS performed the initial literature review, and wrote the first draft under the supervision of SIW. CM and SIW revised the manuscript with support from SF and PS. CM, SF and PS produced the figures. PS and CM performed the calculations. CM and SIW coordinated the research effort.